# Title: Unexpected stable stoichiometries of sodium chlorides

**Authors:** Weiwei Zhang[1, 2, *], Artem R. Oganov[2, 3, 4*], Alexander F. Goncharov[5,6], Qiang Zhu[2], Salah Eddine Boulfelfel[2], Andriy O. Lyakhov[2], Elissaios Stavrou[5], Maddury Somayazulu[5], Vitali B. Prakapenka[7], Zuzana Konôpková[8]

**Affiliations:**
[1] Department of Applied Physics, China Agricultural University, Beijing, 100080, China.
[2] Department of Geosciences, Center for Materials by Design, and Institute for Advanced Computational Science, State University of New York, Stony Brook, NY 11794-2100, U.S.A.
[3] Moscow Institute of Physics and Technology, 9 Institutskiy Lane, Dolgoprudny city, Moscow Region, 141700, Russia
[4] School of Materials Science, Northwestern Polytechnical University, Xi'an,710072, China
[5] Geophysical Laboratory, Carnegie Institution of Washington, 5251 Broad Branch Road NW, Washington, D.C. 20015, U.S.A.
[6] Center for Energy Matter in Extreme Environments and Key Laboratory of Materials Physics, Institute of Solid State Physics, Chinese Academy of Sciences, 350 Shushanghu Road, Hefei, Anhui 230031, China
[7] Center for Advanced Radiation Sources, University of Chicago, Chicago, Illinois 60637, U.S.A.
[8] Photon Science DESY, D-22607 Hamburg, Germany

*To whom correspondence should be addressed.
E-mail: zwwjennifer@gmail.com, artem.oganov@sunysb.edu.

**Abstract:**

**Sodium chloride (NaCl), or rocksalt, is well characterized at ambient pressure. Due to the large electronegativity difference between Na and Cl atoms, it has highly ionic chemical bonding, with stoichiometry 1:1 dictated by charge balance, and B1-type crystal structure. Here, by combining theoretical predictions and diamond anvil cell experiments we show that new materials with different stoichiometries emerge at pressure as low as 20 GPa. Compounds such us $Na_3Cl$, $Na_2Cl$, $Na_3Cl_2$, $NaCl_3$ and $NaCl_7$ are theoretically stable and have unusual bonding and electronic properties. To test this prediction, at 55-80 GPa we synthesized cubic and orthorhombic $NaCl_3$ at 55-70 GPa and 2D-metallic tetragonal $Na_3Cl$. This proves that novel compounds,**

**violating chemical intuition, can be thermodynamically stable even in simplest systems at non-ambient conditions.**

**One Sentence Summary:** Under pressure, sodium chloride loses its iconic simplicity, and textbook chemistry rules break down upon formation of stable compounds $Na_3Cl$, $Na_2Cl$, $Na_3Cl_2$, $NaCl_3$, and $NaCl_7$.

**Main Text:** At ambient conditions, NaCl is the only known stable compound in the Na-Cl system, the chemistry of which is well understood. Extreme conditions, such as high pressure, change the chemical properties of the elements, including Na and Cl (*1, 2*), and the pressure*volume term in the free energy becomes greater than energy of chemical bonds. Thus, strongly compressed matter may exist in totally counterintuitive chemical regimes. Although unexpected high-pressure phenomena may be present in planetary systems or lead to novel exotic materials, we still lack a consistent fundamental understanding of even seemingly simple binary systems like Na-Cl.

The high-pressure behavior of NaCl has been extensively studied experimentally at pressures up to 304 GPa (*3-6*) and by *ab initio* simulations (*7-9*), and very simple behavior was observed - at 30 GPa the rocksalt structure was found to transform into the CsCl (B2-type) structure (*6,8*). From the Herzfeld criterion, metallization of NaCl is expected to occur at 300 GPa (*10*), whereas DFT calculations (*9*) suggest 584 GPa.

To find stable Na-Cl compounds and their structures that may not have been possible to previously observe experimentally or computationally, we used the *ab initio* evolutionary algorithm USPEX (*11-14*), which has a capability to simultaneously find stable stoichiometries and the corresponding structures in multicomponent systems (*15*). In these calculations, any combinations of numbers of atoms in the unit cell were allowed (with the total number ≤16), and calculations were performed at pressures of 1 atm, 20 GPa, 100 GPa, 150 GPa, 200 GPa and 250 GPa. Detailed enthalpy calculations for the most stable structures allowed us to reconstruct the *P-x* phase diagram of the Na-Cl system (Fig.1, Fig.S1-S2).

To verify these predictions, we performed high-pressure experiments in laser heated diamond anvil cell (DAC) at 10-80 GPa on the Na-Cl system in excess of chlorine and sodium (*15*). We specifically targeted synthesis of $Na_3Cl$ and $NaCl_3$, which were predicted to become stable at lowest pressures. The reaction products were examined by visual observations, optical absorption spectroscopy and Raman confocal spectroscopy, and by synchrotron X-ray diffraction probes at room temperature (*15*).

The calculated phase diagram features unexpected compounds - $NaCl_3$, stable above 20 GPa, $NaCl_7$, stable above 142 GPa, and $Na_3Cl_2$, $Na_2Cl$, and $Na_3Cl$, which are stable above 120 GPa, 100 GPa and 77 GPa, respectively. We define as stable those compositions that have lower free energy than any isochemical mixture of other compounds or pure elements. For all the newly predicted structures we computed phonons, and found them to be dynamically stable. In the entire explored pressures region, NaCl is also a stable compound, i.e. will not spontaneously decompose into other compounds. This means that to obtain the newly predicted compounds it is not sufficient just to compress NaCl, but one must do so at high temperatures (to overcome kinetic barriers) and with excess of either Na or Cl. Electronic densities of states (Fig. S5) show that most of these compounds are poor metals, with pronounced pseudogaps at the Fermi level. Cl-rich compounds can be considered as n-type semiconductors, while Na-rich phases are p-type semiconductors. Pseudogaps imply electronic mechanism of their stabilization.

At 20-48 GPa, NaCl$_3$ is stable in the *Pnma* structure, which has 4 formula units in the unit cell. Unlike all the other new phases predicted here (which are metallic), this phase is a semiconductor. Its structure (Fig. 2B) contains almost linear asymmetric Cl$_3$ groups. Bader analysis (*16*) shows that the middle atom in the Cl$_3$-group is nearly neutral, with most negative charge on the side atoms (Table S1) and the total charge of this anion group is ~-0.8. *Pnma*-NaCl$_3$ has ionic bonding between Na$^+$ and [Cl$_3$]$^-$ and rather unusual covalent bonding within [Cl$_3$]$^-$-groups. The latter are reminiscent of well-known trihalide ions I$_3^-$, Br$_3^-$, ClICl$^-$, and of the hypothetical H$_3^-$ ion (*17*), predicted to have charge configuration [H$^{-0.81}$H$^{+0.72}$H$^{-0.81}$]$^{-0.9}$. In a Zintl-like scheme, the central Cl atom must be positively charged to be able to form two covalent bonds and satisfy the octet rule. In the valence-shell electron pair repulsion model (*18*), the central Cl atom of the Cl$_3^-$ group adopts the dsp$^3$ hybridization and has 5 electron pairs, bringing a negative net charge and violating the octet rule. For chlorine atoms, unlike iodine, it is not easy to populate vacant d-orbitals (which nicely explains the structure of ClICl$^-$ ions) and the two schemes work simultaneously – explaining the nearly zero charge of this central atom and its increased (though still relatively small) d-population. At 48 GPa, this peculiar insulating ionic state breaks down, and NaCl$_3$ transforms into a metallic A15 (Cr$_3$Si-type) structure with space group *Pm3n*.

NaCl$_7$, stable above 142 GPa, has cubic structure (space group *Pm*3; Fig. 2A) derivative of the A15 (β-W or Cr$_3$Si) type. The *Pm*3-NaCl$_7$ structure is obtained from *Pm3n*-NaCl$_3$ by substituting the central atom Na (inside Cl$_{12}$-icosahedron) with Cl (Fig. 2A, C). Lattice parameters and bond lengths of $NaCl_3$ and $NaCl_7$ are very close – e.g., within 0.5% at 200 GPa, because at this pressure Na and Cl have almost identical sizes – in stark contrast with ambient conditions, where the $Cl^-$ is much larger than $Na^+$ (ionic radii are 1.81 Å and 1.02 Å, respectively). This opens the possibility of non-stoichiometry and disorder, with the potential for Anderson localization of electronic and phonon states.

At 200 GPa, the shortest Cl-Cl distance in $NaCl_3$ is 2.06 Å, only slightly longer than the Cl-Cl bond in the $Cl_2$ molecule (1.99 Å). These Cl-Cl bonds form extended monatomic chains running along the three mutually perpendicular axes. One recalls a textbook linear chain with a partially filled band – which in the free state is unstable against Peierls distortion (*19*). Isolated Cl-chain has a ½ filled band and at low pressures should break into $Cl_2$ molecules, but in $NaCl_3$ this band is $^2/_3$ filled due to the extra electron donated by Na, and the chain should break into $Cl_3^-$ ions, which we indeed see in the *Pnma*-$NaCl_3$ phase at lower pressures. The application of pressure, and influence of other chemical entities (Na and non-chain Cl atoms), stabilizes these chains in $NaCl_3$, $NaCl_7$, and elemental chlorine. Peierls theorem also explains results of our phonon calculations indicating that both *Pm3n*-$NaCl_3$ and $NaCl_7$ can exist only at high pressure and are not quenchable to atmospheric pressure.

Electronic band structures of $NaCl_3$ and $NaCl_7$ show a deep and wide pseudogap for $NaCl_3$-*Pm3n* (Fig. 3A, B). In both structures, Cl atoms forming the $Cl_{12}$ icosahedra show toroidal ELF maxima (Fig. 3C,D), corresponding to a non-closed-shell electronic configuration, while the Cl1 atom occupying the center of the $Cl_{12}$-icosahedron in $NaCl_7$ has a spherical ELF maximum. Thus, Cl1 and Cl2 atoms in $NaCl_7$ have different electronic structures and play very different chemical roles. Bader analysis (Table S1) confirms this and gives an unusual positive charge to the Cl1.

For the Na-rich side of the phase diagram, we predict several thermodynamically stable compounds – tetragonal $Na_3Cl$ (space group $P4/mmm$), two phases of $Na_3Cl_2$: tetragonal (space group $P4/m$) and orthorhombic (space group $Cmmm$), and three phases of $Na_2Cl$ – one tetragonal (space group $P4/mmm$), and two orthorhombic (space groups $Cmmm$ and $Imma$). Most of these are layered superstructures of the CsCl-type (B2) structure, with both Na and Cl atoms in the eightfold coordination (Fig. 2). For example, $Na_3Cl$ can be represented as a $[NaCl][Na_2][NaCl][Na_2]$… sequence of layers, and the *c*-parameter of the unit cell is doubled relative to that of B2-NaCl. This and similar structures have very interesting 2D-metallic features, with alternating metallic $[Na_2]$ and insulating $[NaCl]$ layers. $Na_3Cl_2$ is stable above 120 GPa, and its $P4/m$ structure can be described as a 1D- (rather than layered, 2D-) ordered substitutional superstructure of the B2-NaCl structure. Compound $Na_2Cl$ shows a more complex behavior than $Na_3Cl$ or $Na_3Cl_2$. At 100-135 GPa, the $P4/mmm$ structure is stable; it is also a layered B2-type superstructure. At 135-298 GPa, $Na_2Cl$ is stable in the $Cmmm$ structure, and above 298 GPa in the $Imma$ structure with Na and Cl atoms in the 12- and 10-fold coordination, respectively (Fig. 2).

XRD measurements of $Na_3Cl$ and $NaCl_3$ synthesized at high pressures show new Bragg peaks after laser heating. At pressures above 60 GPa these Bragg peaks can be indexed either in a cubic $NaCl_3$ unit cell (Fig. 4A) or with a mixture of the cubic and the orthorhombic $Pnma$ $NaCl_3$ unit cell. With pressure decreasing below 54 GPa, after laser heating, only the peaks of the orthorhombic $NaCl_3$ are present in the XRD patterns (Fig.S7). The XRD pattern usually also contains peaks from unreacted B2-NaCl and orthorhombic chlorine, which were identified using our theoretical calculations.

From the XRD data, we obtained the lattice parameters and the unit cell volume for the two structures as a function of pressure for the decompression sequence. There is good agreement between the experimental and theoretical equations of state for both $NaCl_3$ structures (Fig. 4C). Also, in agreement with the theoretical predictions, we find that two new $NaCl_3$ phases transform from one to another upon pressure release at 300 K, although the transition is sluggish and there is a large range of phase coexistence. *Pnma*-$NaCl_3$ remains metastable at pressures as low as 18 GPa, and decomposes to NaCl and $Cl_2$ at lower pressures.

Raman spectroscopy of reacted Cl-rich compounds confirms the XRD data. We have observed Raman spectra of two different kinds, depending on pressure and observation points, they were even superimposed. The Raman spectra of the excess $Cl_2$ were easy to identify because we also collected correspond to reference data for unreacted materials as a function of pressure. The Raman spectra of two polymorphic modifications of $NaCl_3$ (Fig.S8, S9) are quite different. In cubic $NaCl_3$ we observed one broad strong band near 450 $cm^{-1}$ and a number of weaker features, while in *Pnma* $NaCl_3$ there are a number of narrow peaks. In both cases the agreement between theory and experiment is good concerning the major peak positions. Moreover, experimental and theoretical pressure dependences of the Raman frequencies (Fig S10) are also in good agreement for both structures. However, in cubic phase we observed a number of extra peaks (Raman forbidden for *Pm3n* lattice) corresponding to other zone-center phonons, *i.e.* some of the selection rules appear to be lifted. These selection rules could be substantially lifted in surface Raman scattering (as happens in metals), due to disorder in site occupation or even a variable stochiometry as Cl and Na are easily interchangeable at high pressure. Optical absorption spectra of the synthesized material (Fig.S6) show a gap-like feature at 1.7 eV, which is consistent with the predicted prominent pseudogap in the electronic density of states (Fig. 3B).

The case of Na-rich material is similar, but less complex. XRD of the samples laser heated above 60 GPa shows new Bragg peaks that can be indexed in a tetragonal *P4/mmm* unit cell (Fig. 4B) across the whole pressure range of this study. The XRD pattern usually also contains peaks from unreacted cubic B1 or B2 NaCl and bcc or fcc sodium (*20*). The lattice parameters of new material agree well with the theoretically predicted *P4/mmm* $Na_3Cl$ in a wide pressure range of 27-70 GPa (Fig. 4D). The Raman data (Fig. S11) are also consistent with theoretical predictions. However, as in the case of the Cl-rich compounds, in addition to two Raman-allowed modes, we also observe a number of Raman forbidden bands, whose positions agree reasonably well with other computed zone-center phonons. Finally, Raman data show that newly synthesized Na-rich material can be (meta)stable down to 20 GPa (Fig.S11,S12) and then decomposes into NaCl and Na at lower pressures.

The theoretical prediction, and experimental synthesis, of unexpected chemical compounds in a simple binary system, such as Na-Cl, is not entirely unexpected. Counterintuitive compounds (such as $LiH_2$, $LiH_6$, $LiH_8$) have been predicted (*21*) to appear under pressure, but experiments failed to find them so far (*22*). Our results suggest that new stable compositions with unusual chemical bonding may exist in other simple systems, such as K-Cl, but also in important planet-forming systems like Mg-Si-O (*23*) and H-C-N-O. Furthermore, these results point to possibilities for creating materials with unusual properties that may be quenchable to ambient conditions.

**Acknowledgments:** We thank the National Science Foundation (EAR-1114313, DMR-1231586), DARPA (Grants W31P4Q1310005 and W31P4Q1210008), the Government of the Russian Federation (grant #14.A12.31.0003), China's Foreign Talents Introduction and Academic Exchange Program (#B08040), and German BMBF (#05K10RFA) for financial support. W.W.Z acknowledges support from the Young Teachers Development Project in China Agricultural University. A.F.G. acknowledges support from the NSF, Army Research Office, and EFREE, a BES-EFRC center at Carnegie. Calculations were performed on XSEDE facilities (Charge #TG-DMR110058) and on the cluster of the Center for Functional Nanomaterials, Brookhaven National Laboratory, which is supported by the DOE-BES under contract DE-AC02-98CH10086. X-ray diffraction experiments were performed at GeoSoilEnviroCARS (Sector 13), Advanced Photon Source (APS), Argonne National Laboratory and Petra III, DESY, Hamburg, Germany. GeoSoilEnviroCARS is supported by the National Science Foundation - Earth Sciences (EAR-1128799) and Department of Energy - Geosciences (DE-FG02-94ER14466). Use of the Advanced Photon Source was supported by the U. S. Department of Energy, Office of Science, Office of Basic Energy Sciences, under Contract DE-AC02-06CH11357. PETRA III at DESY is a member of the Helmholtz Association (HGF). USPEX code is available at http://uspex.stonybrook.edu and experimental data are available in the Supplementary Materials.

Author contributions: A.R.O. designed the research. W.W.Z., Q.Z., S.E.B. and A.R.O. performed the calculations, interpreted data and wrote the paper. A.L. wrote the latest version of the structure prediction code. A.F.G., E.S. performed the experiments, interpreted the data and contributed in writing the manuscript. M.S., V.P., and Z.K. performed the experiments and contributed into the experimental methods. W.W.Z. and A.R.O contributed equally to this paper.

**Supplementary Materials**

Materials and Methods
Supplementary Text
Figs. S1 to S12
Table S1
References (11-14, 24-32)

**Fig. 1. Stability of new sodium chlorides:** (A) Pressure-composition phase diagram of the Na-Cl system. (B) Convex hull diagram for Na-Cl system at selected pressures. Solid circles represent stable compounds; open circles - metastable compounds.

**Fig. 2. Crystal structures of Na chlorides and NaCl$_7$:** (A) *Pm3*-NaCl$_7$, (B) *Pnma*-NaCl$_3$, (C) *Pm3n*-NaCl$_3$, (D) *P4/mmm*-Na$_3$Cl, (E) *P4/m*-Na$_3$Cl$_2$, (F) *Cmmm*-Na$_3$Cl$_2$, (G) *P4/mmm*-Na$_2$Cl, (H) *Cmmm*-Na$_2$Cl, and (I) *Imma*-Na$_2$Cl. Blue and green spheres – Na and Cl atoms, respectively.

**Fig. 3. Electronic structure of NaCl$_7$ and NaCl$_3$ at 200 GPa.** Band structure and density of states of (A) NaCl$_7$ and (B) NaCl$_3$ and electron localization function of (C) NaCl$_7$ and (D) NaCl$_3$ at ELF= 0.80. For clarity, atom-projected DOSs in (A, B) were multiplied by 3 (for NaCl$_7$) and by 4 (for NaCl$_3$).

**Fig. 4. Powder X-ray diffraction patterns and equations of state for NaCl$_3$ and Na$_3$Cl.** (A) NaCl$_3$ in Cl$_2$ medium and (B) Na$_3$Cl in Na medium collected at 60 GPa. Vertical ticks correspond to Bragg peaks of Na-Cl phases. The X-ray wavelengths are 0.5146 Å in (A) and 0.3344 Å in (B). The peak marked by the asterisk corresponds to the strongest peak of rhenium (gasket material). Equations of state of (C) NaCl$_3$ and (D) Na$_3$Cl as determined experimentally (symbols) and theoretically (lines).

A

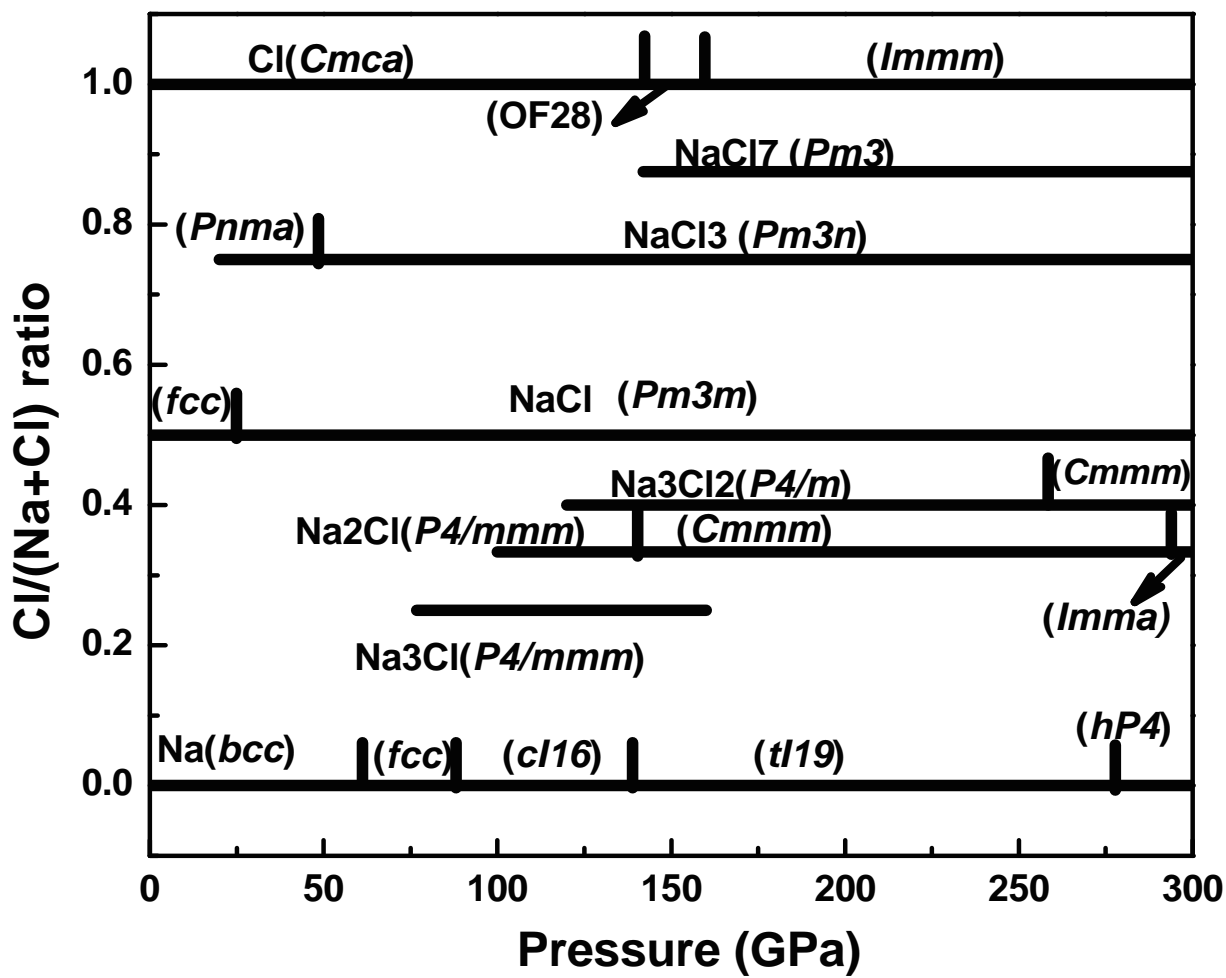

B

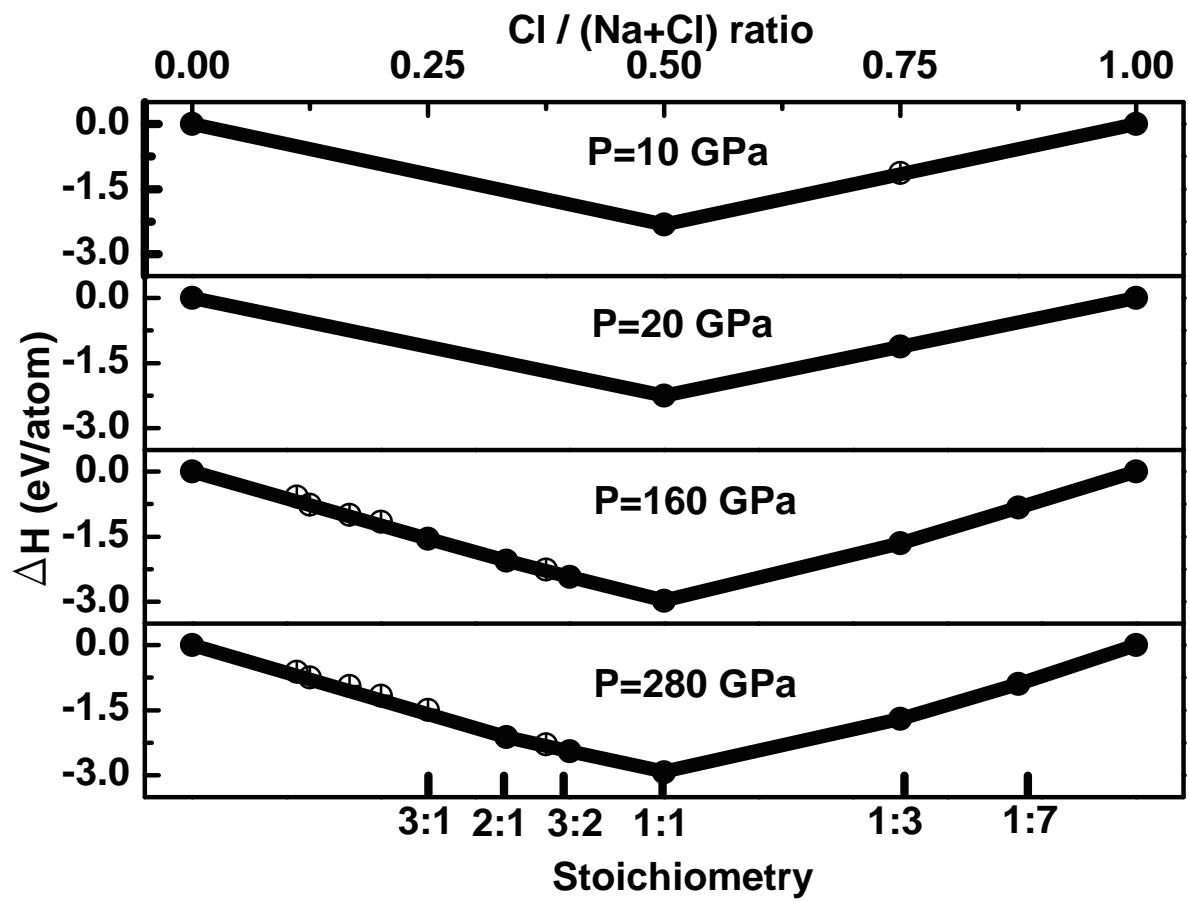

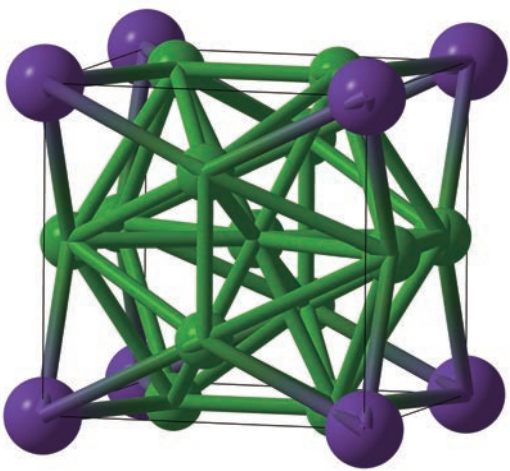

A

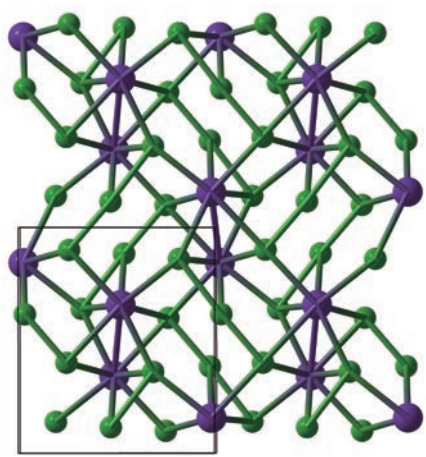

B

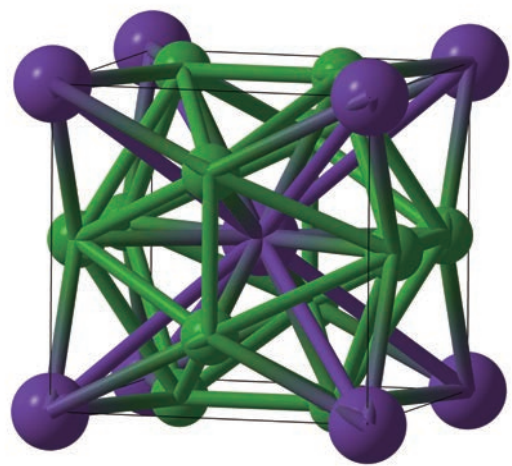

C

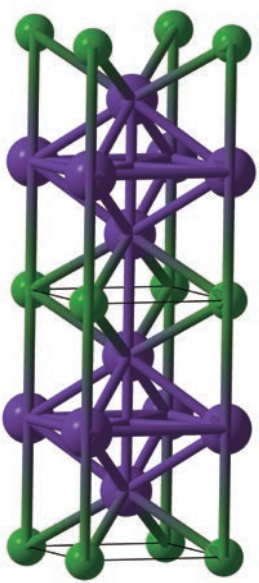

D

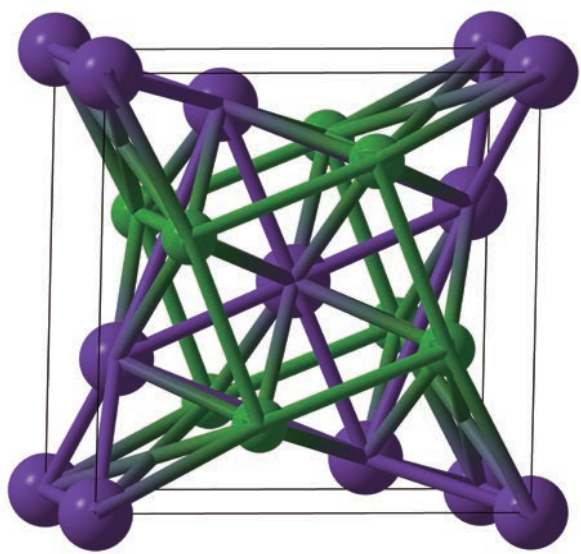

E

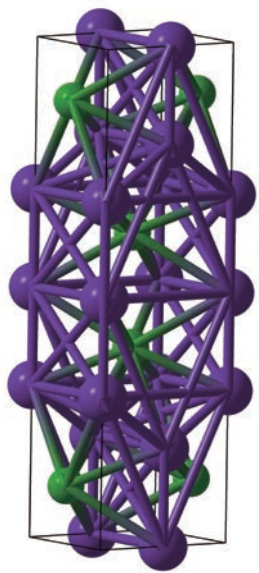

F

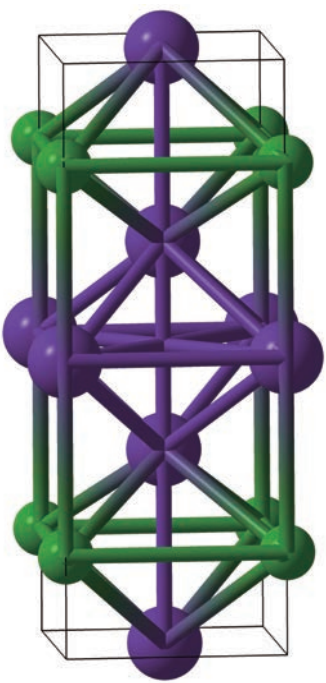

G

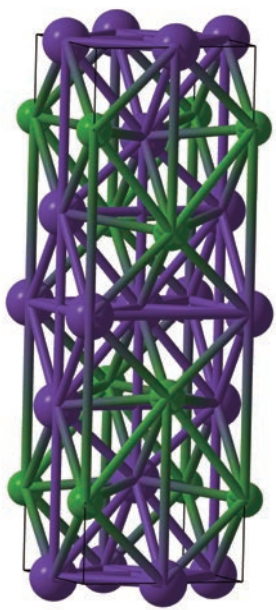

H

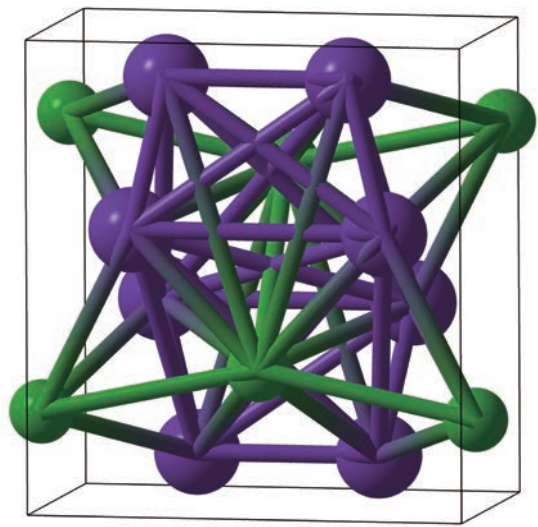

I

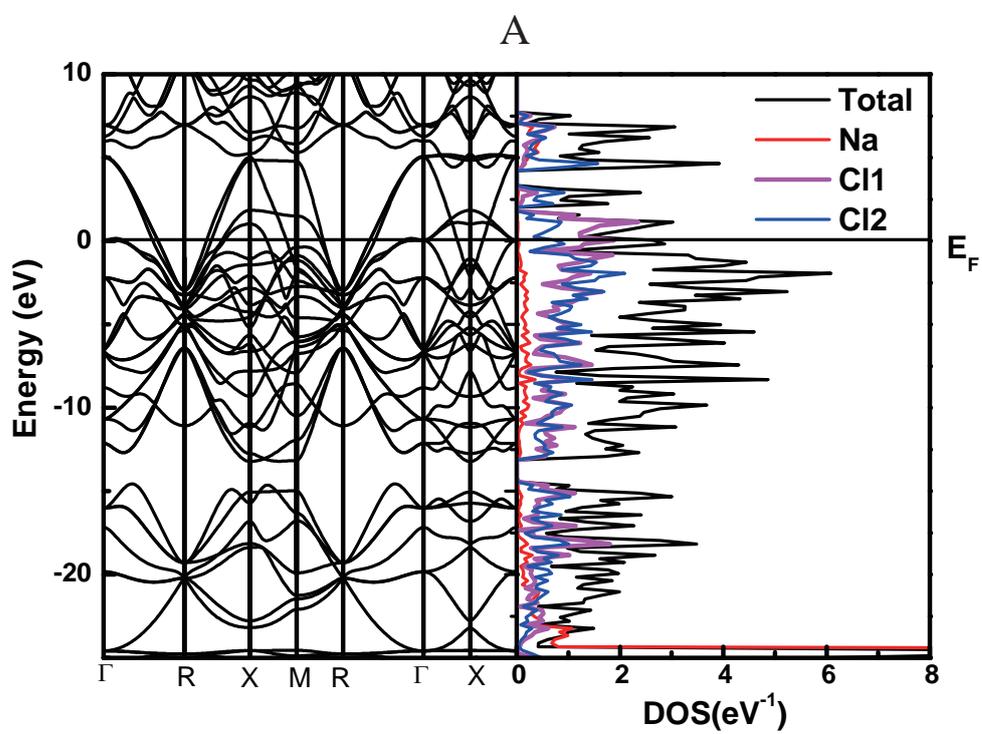

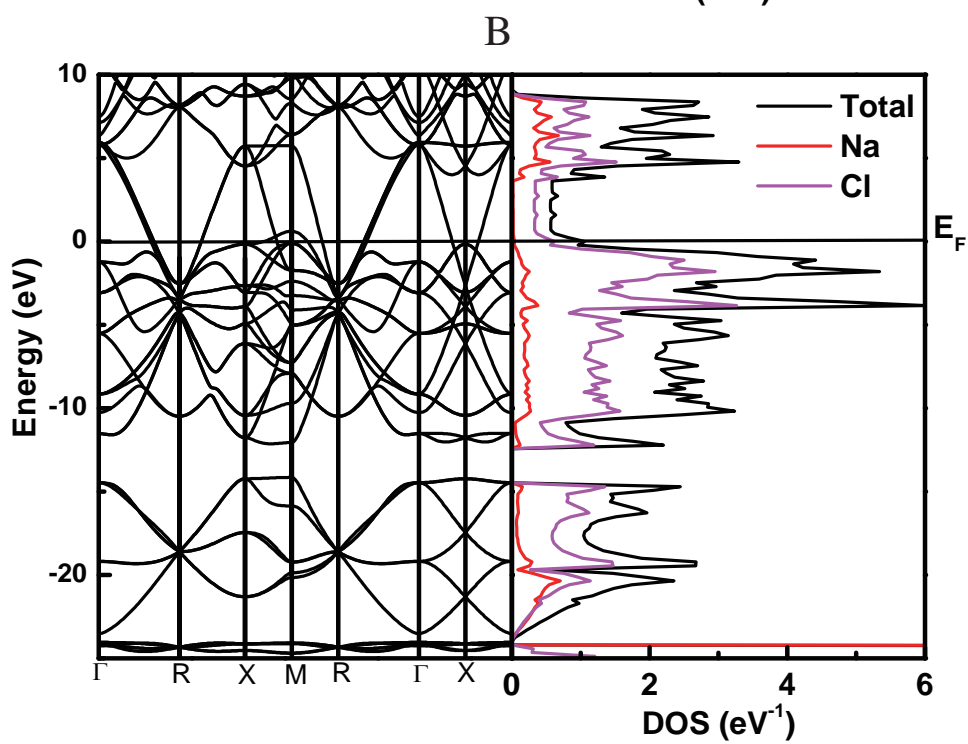

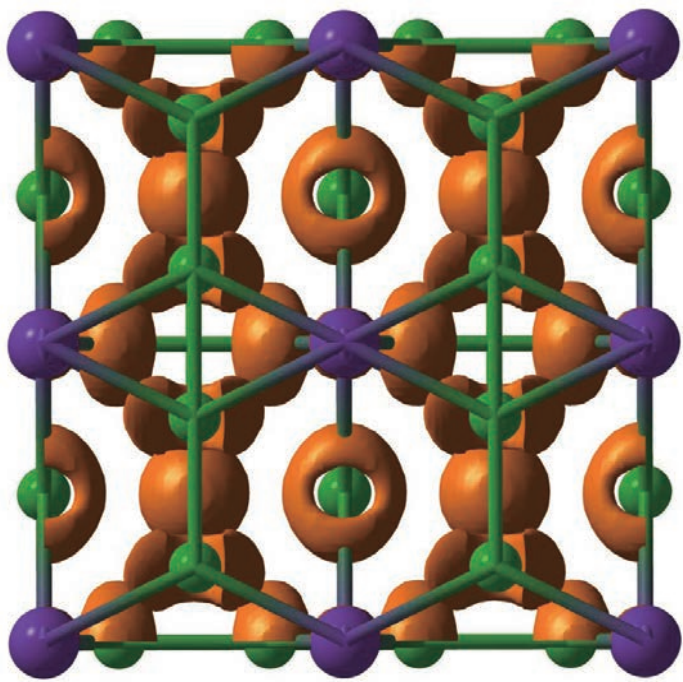
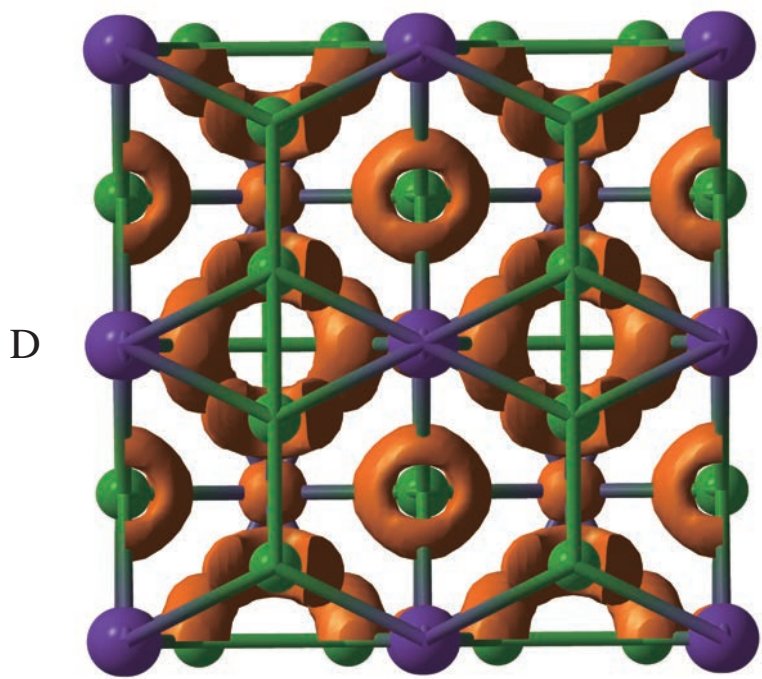

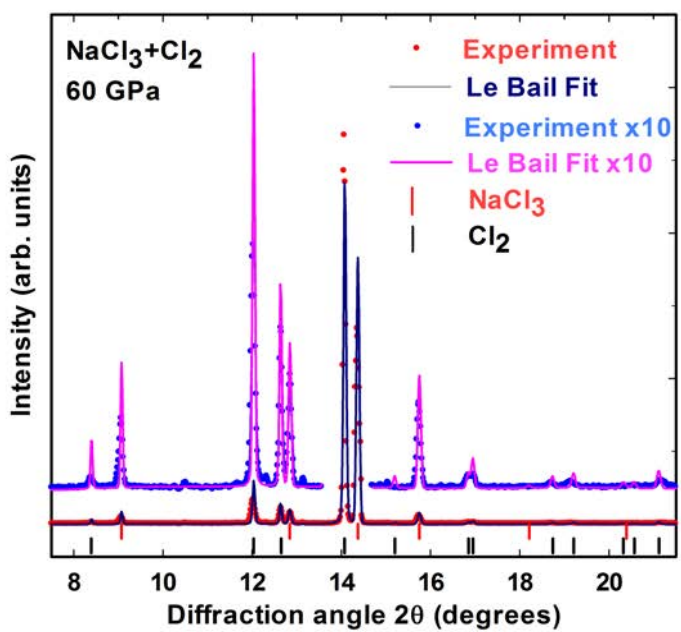
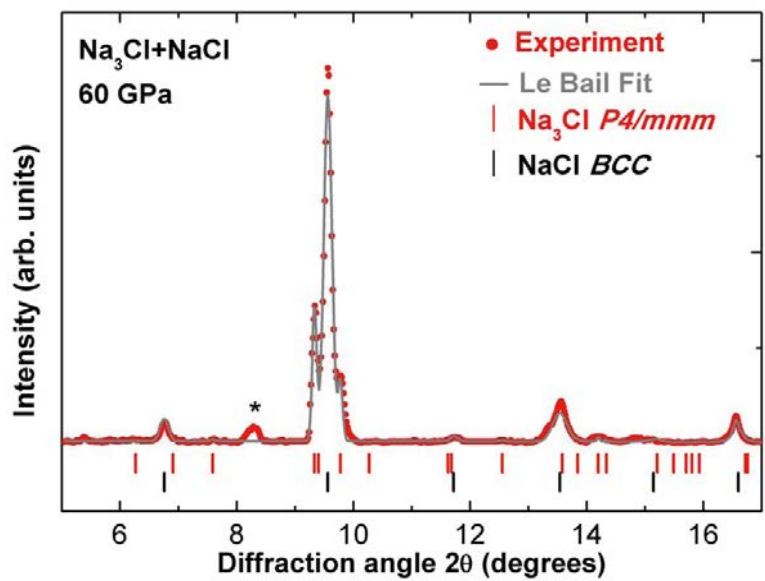
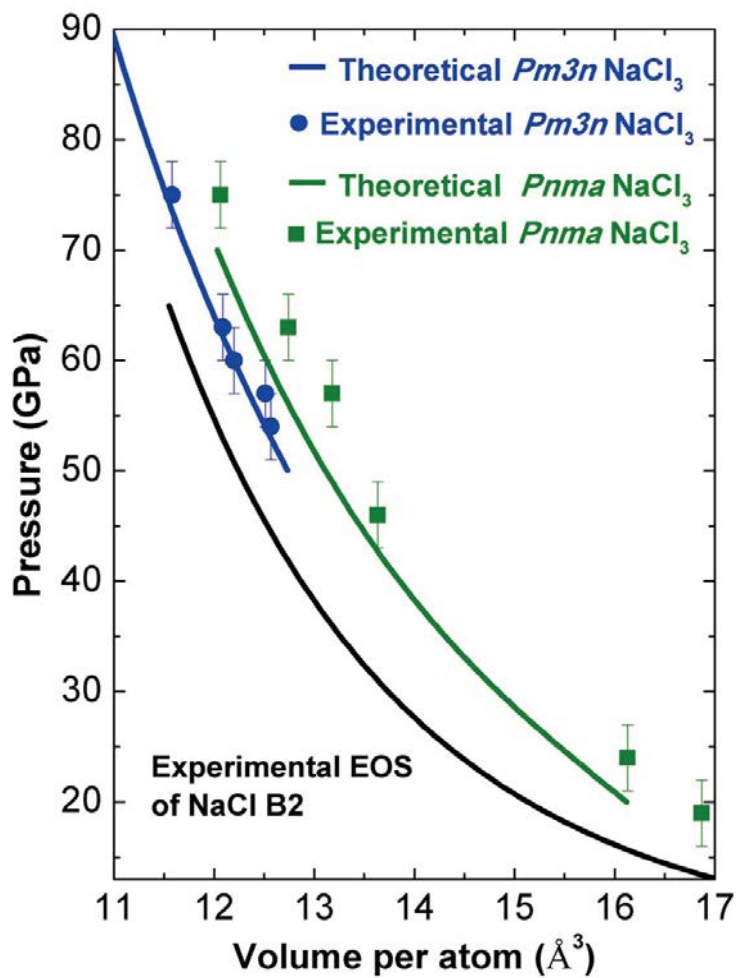
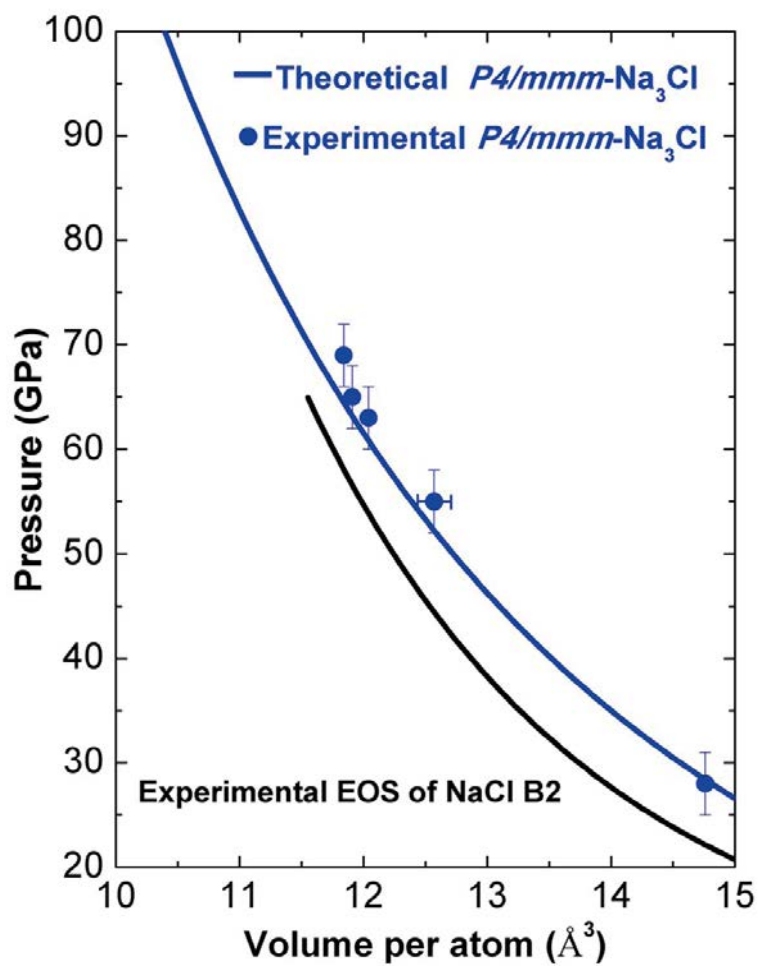

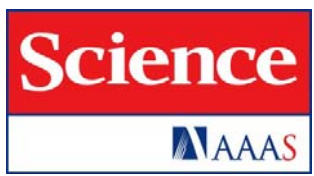

# Supplementary Materials for

## Unexpected stable stoichiometries of sodium chlorides

Weiwei Zhang, Artem R. Oganov, Alexander F. Goncharov, Qiang Zhu, Salah Eddine

Boulfelfel , Andriy O. Lyakhov, Elissaios Stavrou, Maddury Somayazulu, Vitali

Prakapenka, Zuzana Konopkova

Correspondence to:   zwwjennifer@gmail.com, artem.oganov@sunysb.edu

**This PDF file includes:**

Materials and Methods
Supplementary Text
Figs. S1 to S12
Table S1
References (11-14, 24-32)

**Other Suplementary Materials for this manuscript includes the following:**
Typical input file for USPEX code

# Supplementary Materials for

**Unexpected stable stoichiometries of sodium chlorides**

**Authors:** Weiwei Zhang[1, 2, *], Artem R. Oganov[2, 3, 4*], Alexander F. Goncharov[5,6], Qiang Zhu[2], Salah Eddine Boulfelfel[2], Andriy O. Lyakhov[2], Elissaios Stavrou[5], Maddury Somayazulu[5], Vitali B. Prakapenka[7], Zuzana Konôpková[8]

[1] Department of Applied Physics, China Agricultural University, Beijing, 100080, China. [2] Department of Geosciences, Center for Materials by Design, and Institute for Advanced Computational Science, State University of New York, Stony Brook, NY 11794-2100, U.S.A. [3] Moscow Institute of Physics and Technology, 9 Institutskiy Lane, Dolgoprudny city, Moscow Region, 141700, Russia. [4] School of Materials Science, Northwestern Polytechnical University, Xi'an, 710072, China. [5] Geophysical Laboratory, Carnegie Institution of Washington, 5251 Broad Branch Road NW, Washington, D.C. 20015, U.S.A. [6] Center for Energy Matter in Extreme Environments and Key Laboratory of Materials Physics, Institute of Solid State Physics, Chinese Academy of Sciences, 350 Shushanghu Road, Hefei, Anhui 230031, China. [7] Center for Advanced Radiation Sources, University of Chicago, Chicago, Illinois 60637, U.S.A. [8] Photon Science DESY, D-22607 Hamburg, Germany

*To whom correspondence should be addressed.
E-mail: zwwjennifer@gmail.com, artem.oganov@sunysb.edu.

−*These authors contributed equally to this work.

## Materials and Methods

Calculations.

Predictions of stable phases were done using the USPEX code (*11-13*) in the variable-composition mode (*14*). The first generation of structures was produced randomly and the succeeding generations were obtained by applying heredity, atom transmutation, and lattice mutation operations, with probabilities of 60%, 10% and 30%, respectively. 80% non-identical structures of each generation with the lowest enthalpies were used to produce the next generation. All structures were relaxed using density functional theory (DFT) calculations within the Perdew-Burke-Ernzerhof (PBE) (*24*), as implemented in the VASP code (*25*). We used the all-electron projector augmented wave (PAW) (*26*) with [He] core (radius 1.45 a.u.) for Na and [Ne] core (radius 1.50 a.u.) for Cl, plane-wave basis sets with the 980 eV cutoff, and dense Monkhorst-Pack meshes with resolution $2\pi \times 0.05$ Å$^{-1}$. Having identified the most stable compositions and structures, we relaxed them at pressures from 1 atm to 300 GPa with an even denser Monkhorst-Pack mesh with resolution $2\pi \times 0.03$Å$^{-1}$. Bader charge analysis was done using the grid-based algorithm (*27*) with 120x120x120 grids.

Phonons and Raman spectra were computed within density-functional perturbation theory as implemented in the Quantum Espresso package (*28*). We used a plane-wave kinetic energy cutoff of 180 Ry, in conjunction with dense k- and q-point meshes. For example, for *Pm*3*n*-NaCl$_3$, we used a 12×12×12 mesh for the Brillouin zone, and 4×4×4 *q*-mesh for computing the force constants matrix. The electron-phonon coupling matrix elements for the electron-phonon interaction coefficients were calculated with a large 16×16×16 grid.

We find that phase stability fields in Fig. 1 (computed without zero-point vibrational energies) are only slightly affected by the zero-point energy. For example, the reaction NaCl + 2Na = Na$_3$Cl is predicted to be thermodynamically favorable above 77 GPa if zero-point energy is neglected, and above 82 GPa when it is included.

Experiments.

We loaded two stacked NaCl thin (5-8 μm) plates of 50 x 50 μm dimensions in the DAC cavity of 80 μm diameter made in preindented to 35 μm thickness rhenium gasket and filled the rest of the cavity either cryogenically by molecular chlorine or with sodium inside a glove box for the Cl and Na rich compounds, respectively. Diamond anvils with 200 μm diameter flat tips were used. Optical absorption and Raman spectra were monitored on pressure increase. At 55-60 GPa due to the bandgap narrowing, Cl$_2$ became sufficiently absorptive to couple to a 1075 nm fiber laser radiation. Laser heating at 55-80 GPa via absorption by chlorine (which reduces sufficiently its semiconducting gap under pressure) or metallic Na results in a chemical reaction, which was detected by a sudden increase in temperature near and above 2000 K. For both setups (excess of chlorine and excess of sodium), experiments were repeated several times and with different starting pressures *i.e.* the pressure where the laser heating has been performed.

We laser heated the NaCl plate which is insulated from the diamond anvils from all sides by chlorine, as this configuration allows heating it from both sides. In the case of excess of Na, coupling was achieved through the metallic Na. In this case, we loaded in a glove box a small piece of Na in the DAC cavity between two thin NaCl plates positioned on each diamond anvil.

The laser heating remains very local during this procedure as our radiometric measurements and finite element calculations show. Thus, we do not expect any reaction with a gasket material (which remains cold during the heating) or with diamond anvils; this was verified by reversibility in pressure of the Raman observations.

Raman studies were performed using 488 and 532 nm lines of a solid-state laser. The laser probing spot dimension was 4 μm. Raman spectra were analyzed with a spectral resolution of 4 cm$^{-1}$ using a single-stage grating spectrograph equipped with a CCD array detector. Optical absorption spectra in visible and near IR spectral ranges were measured using an all-mirror custom microscope system coupled to a grating spectrometer equipped with a CCD detector (*29*). Laser heating was performed in a double-sided laser heating system combined with a confocal Raman probe (*30, 31*) and also at the undulator XRD beamline at GeoSoilEnviroCARS, APS, Chicago and Extreme Conditions Beamline P02.2 at DESY (Germany), which have online laser heating capabilities. Temperature was determined spectroradiometrically. Synchrotron XRD data were also collected on the quenched samples using bending magnet beamlines of GeoSoilEnviroCARS and of HPCAT at the Advanced Photon Source (*32*). The X-ray probing beam size was about 10 μm at the bending beamlines, and 2-5 μm at the undulator beamlines.

**Supplementary text**

**<u>Brief descriptions of the crystal structures.</u>**

   **NaCl$_7$** is stable above 142 GPa and has a cubic structure (space group *Pm*3) with 1 formula unit (f.u.) in the primitive cell. At 200 GPa it has the optimized lattice parameter *a* = 4.133 Å, with Na atoms occupying the Wyckoff 1a (0.0, 0.0, 0.0) position; there are two inequivalent Cl sites – Cl1 atoms occupy the 1b (0.5, 0.5, 0.5) and Cl2 the 6g (x, 0.5, 0.0) site with x=0.248. Na atoms sit at the corners of the cubic unit cell, while Cl2 atoms form an icosahedron around the Cl1 atoms. The nearest distance between Cl1 and Cl2 is 2.31 Å, whereas the shortest Cl2-Cl2 distance is 2.05 Å. This structure is shown in Fig. 2(b) and can be described as a derivative of the A15 (β-W or Cr$_3$Si) structure type.

   **NaCl$_3$-*Pnma*** structure is stable at 20-48 GPa. It has 4 f.u. in the unit cell. This structure at 40 GPa has parameters *a* = 7.497 Å, *b* = 4.539 Å, *c* = 6.510 Å, with all atoms occupying Wyckoff 4c (x, 0.25, z) sites with x=0.342 and z= 0.480 for Na, x = 0.141 and z = 0.180 for Cl1, x = 0.095 and z = 0.741 for Cl2, and x = 0.383 and z = 0.966 for Cl3.

   **NaCl$_3$–*Pm*3*n*** is predicted to become stable above 48 GPa. This is a metal with an A15 (Cr$_3$Si-type) structure with space group *Pm*3*n* and 2 f.u. in the primitive cell. At 200 GPa, the *Pm*3*n*-NaCl$_3$ structure has the optimized lattice parameter *a* = 4.114 Å with Na and Cl occupying the Wyckoff 2a (0.0, 0.0, 0.0) and 6d (0.25, 0.5, 0.0) positions, respectively. The nearest Na-Cl distance is 2.30 Å, whereas the shortest Cl-Cl distance is 2.06 Å - only slightly longer than the bond length in the Cl$_2$ molecule (1.99 Å). However, here and in NaCl$_7$, there are no molecules and short Cl-Cl bonds form extended monatomic chains running along the three mutually perpendicular axes. The A15 structure is common for type-II high-*T*$_c$ superconductors (which before the discovery of cuprate superconductors for decades held the record of highest known *T*$_C$ values) with strong electron-phonon coupling. However, our calculations for A15-type NaCl$_3$ found no evidence of superconductivity.

   ***P*4/*mmm*-Na$_3$Cl** at 140 GPa has lattice parameters *a* = 2.786 Å and *c* =4.811 Å, with Na occupying the Wyckoff 1b (0.0, 0.0, 0.5) and 2h (0.5, 0.5, 0.238) positions, and Cl atoms at 1a (0.0, 0.0, 0.0) sites.

   ***Pm*3*m*-NaCl** at 140 GPa has lattice parameter *a* =2.667 Å, and Na and Cl occupy the 1a (0.0, 0.0, 0.0) and 1b (0.5, 0.5, 0.5) sites, respectively.

   ***P*4/*m*-Na$_3$Cl$_2$** structure at 140 GPa has lattice parameters *a* =5.753 Å and *c* =2.820 Å, with Na atoms occupying the Wyckoff 4k (0.901, 0.676, 0.5), 1d (0.5, 0.5, 0.5), and 1a (0.0, 0.0, 0.0) positions, and Cl at 4j (0.799, 0.372, 0.0) sites.

   ***Cmmm*-Na$_3$Cl$_2$** structure at 280 GPa has lattice parameters *a*=9.881 Å, *b*=2.925 Å and c=2.508 Å with Na atoms occupying the Wyckoff 2b (0.5, 0.0, 0.0), 4g (x, 0.5, 0) positions with x= 0.196 and Cl 4h (x, 0.0, 0.5) at sites with x=0.115.

**P4/mmm-Na₂Cl** at 120 GPa has lattice parameters $a$ =2.790 Å and $c$ =7.569 Å, with Na atoms occupying the Wyckoff 1b (0.0, 0.0, 0.5), 1c (0.5, 0.5, 0.0), and 2g (0.0, 0.0, 0.827) positions, and Cl sitting at 2h (0.5, 0.5, 0.673) sites.

**Cmmm-Na₂Cl** at 180 GPa has lattice parameters $a$ = 3.291 Å, $b$ =10.385 Å, $c$ =2.984 Å, and Na atoms at the Wyckoff 2b (0.0, 0.5, 0.0), 2d (0.5, 0.5, 0.5), and 4j (0.5, 0.181, 0.5) positions, and Cl in the 4i (0.0, 0.153, 0.0) sites.

**Imma-Na₂Cl** at 300 GPa has lattice parameters $a$ = 4.929 Å, $b$ =3.106 Å, $c$ =5.353 Å, with Na and Cl atoms occupying the Wyckoff 8i (0.791, 0.75, 0.588) and 4e (0.0, 0.25, 0.798) positions, respectively.

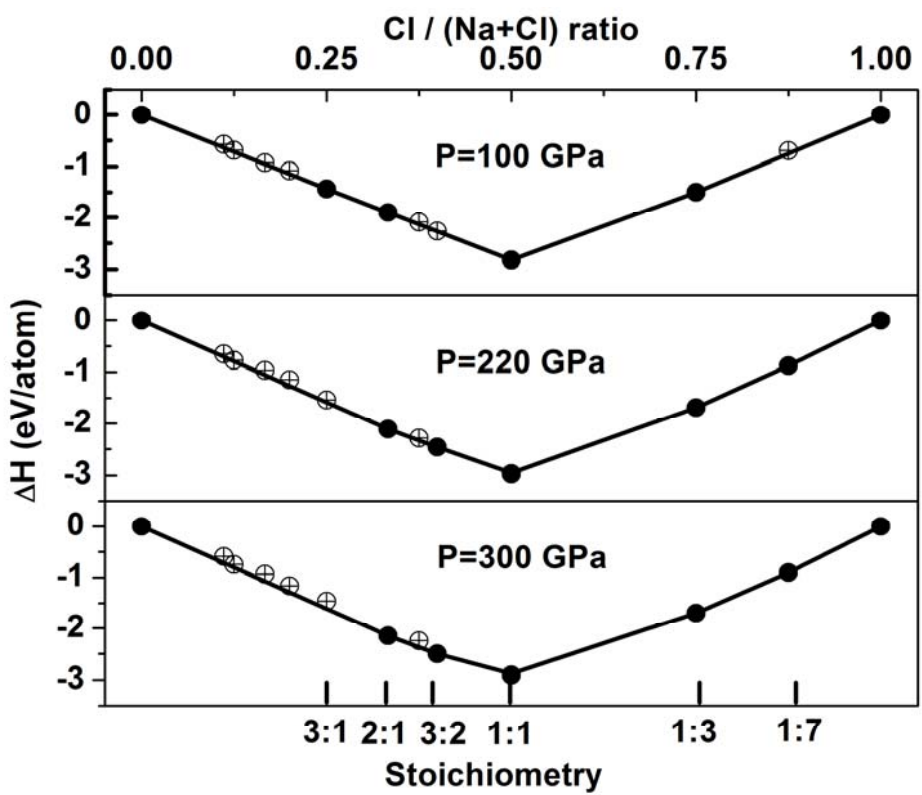

Figure S1. Convex hull diagrams for the Na-Cl system at several pressures. Solid circles represent stable structures; open circles metastable structures.

A

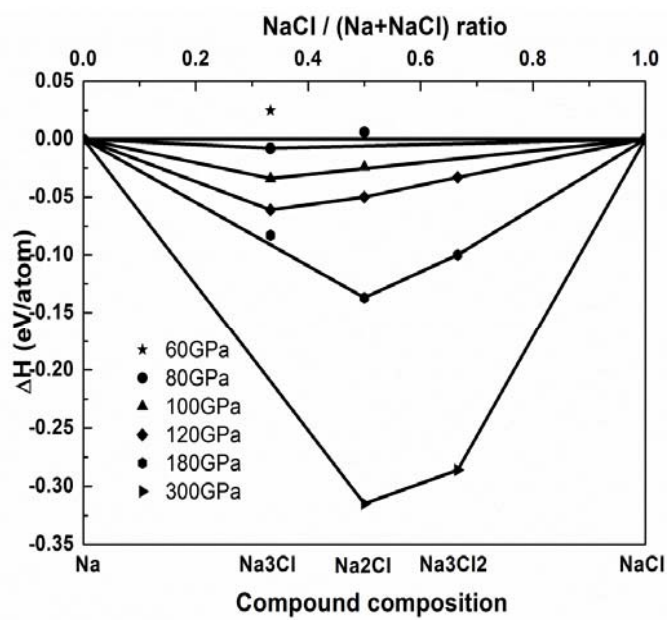

B

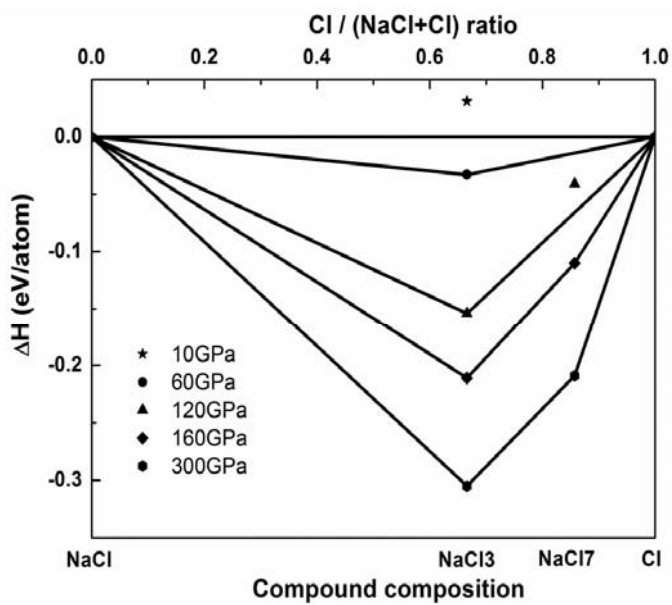

Figure S2. Convex hull diagrams for the (A) Na-NaCl and (B) NaCl-Cl systems. This illustrates that the formation of new sodium chlorides is a strongly exothermic process.

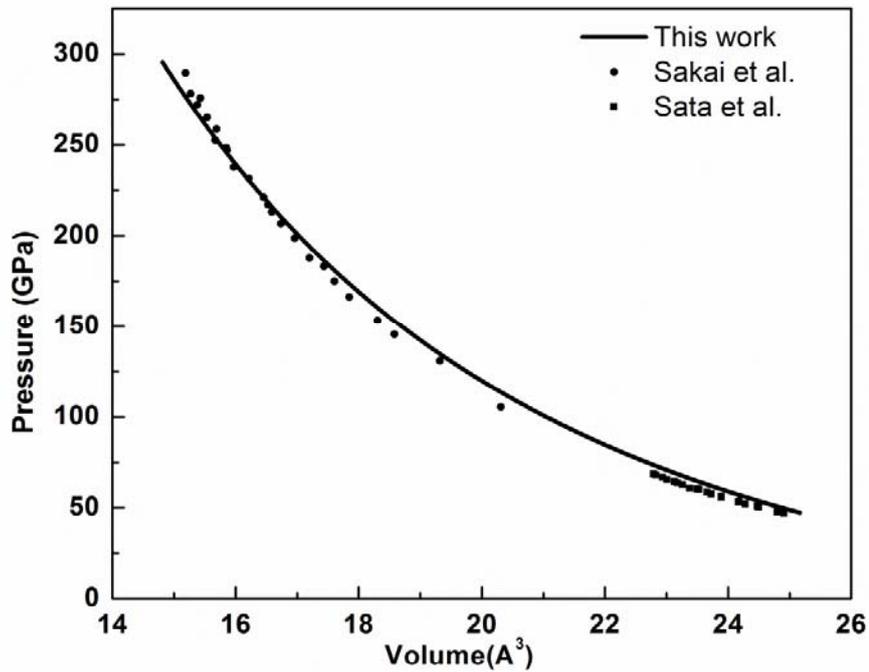

Figure S3. Equation of state of B2-NaCl from experiment (*4, 5*) and theory (this work). The solid line is a $3^{rd}$-order Birch-Murnaghan fit to the energy-volume data from 40 GPa to 300 GPa. Circles are from experiments of Ref.5 with pressure determined using the equation of state of platinum. Squares are from experiments of Ref.4 with pressure determined using the equation of state of MgO. This simple comparison shows how accurately theoretical calculations model high-pressure behavior of NaCl.

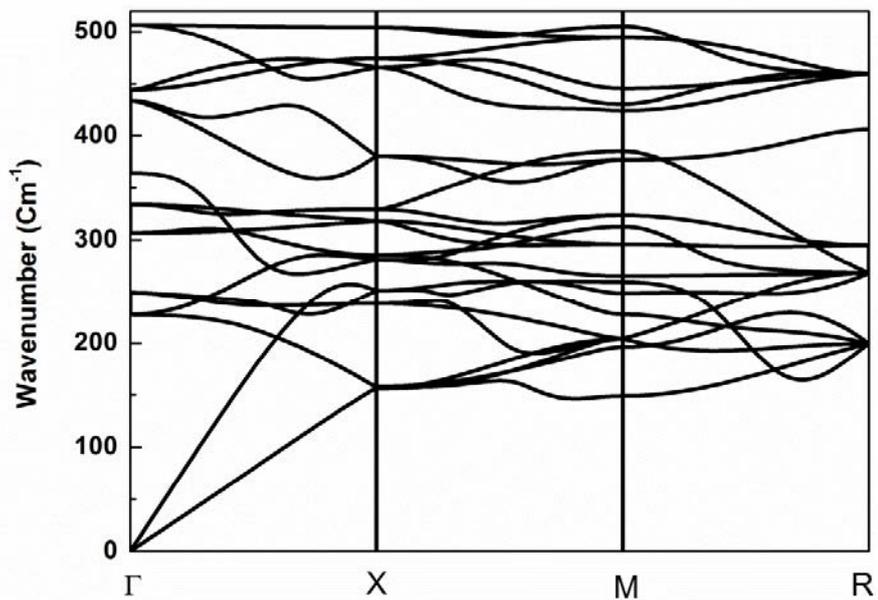

Figure S4. Phonon dispersion curves of cubic $NaCl_3$ at 60 GPa. Such calculations were done for all predicted structures to ensure their dynamical stability.

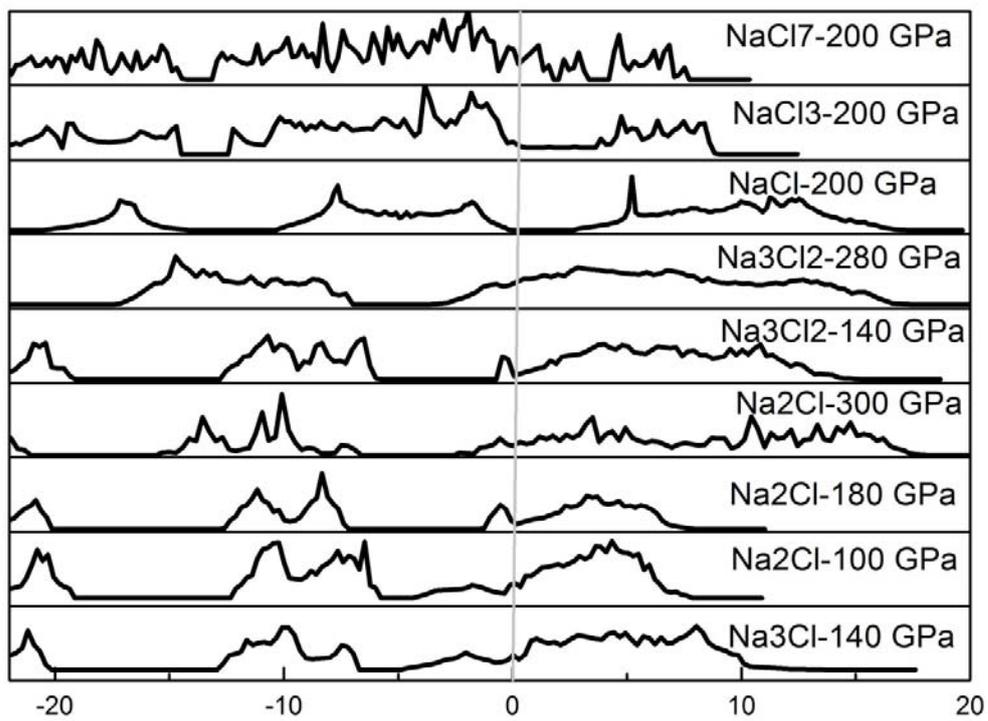

Figure S5. Electronic densities of states of sodium chlorides.

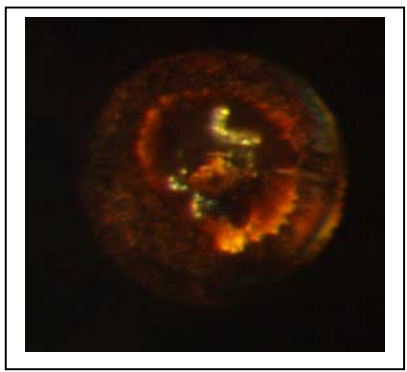
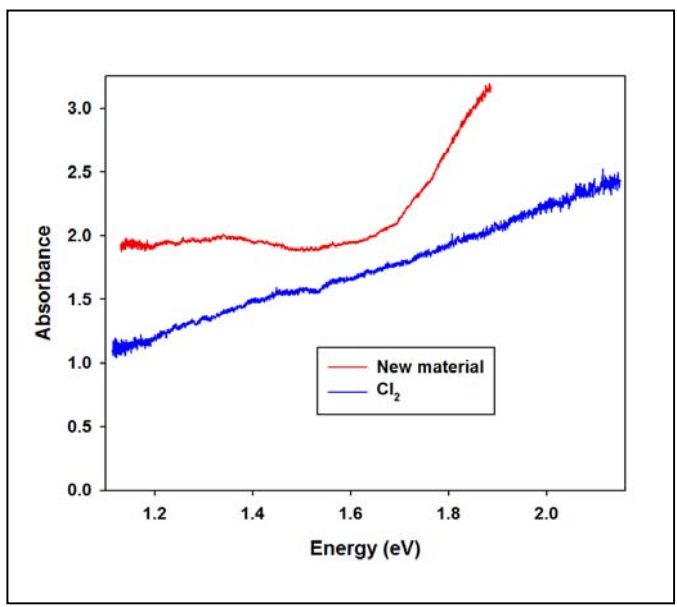
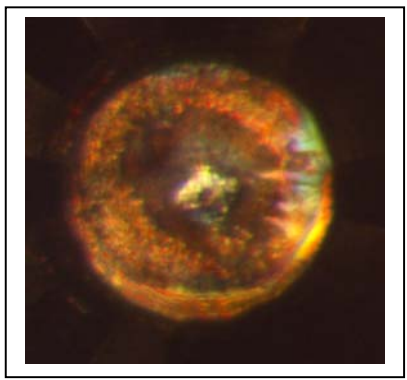

Figure S6. Optical properties of a material synthesized at 60 GPa and 2000 K in the DAC. (A) Optical photographs of the sample in transmitted and reflected light (top) and only reflected (bottom) light, demonstrating strong reflectivity. (B) Optical absorption spectra of a new material in comparison to that of $Cl_2$ medium. The former one can have a background contribution from unreacted chlorine medium. The latter we monitored as a function of pressure and showed a monotonous red shift without a substantial change in shape.

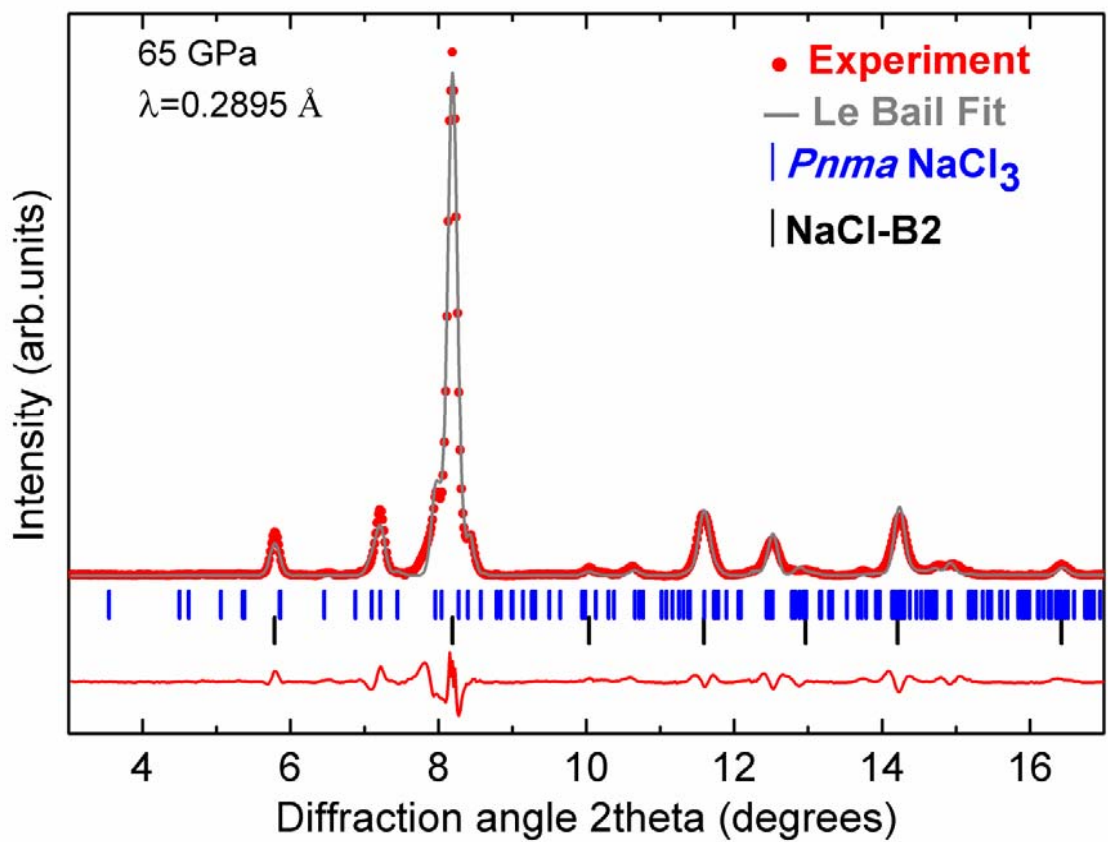

Figure S7. Powder X-ray diffraction patterns of $NaCl_3$ at 65 GPa. Vertical ticks correspond to Bragg peaks of *Pnma*-$NaCl_3$ (blue) and $B_2$-NaCl (black). The X-ray wavelength is 0.2895 Å

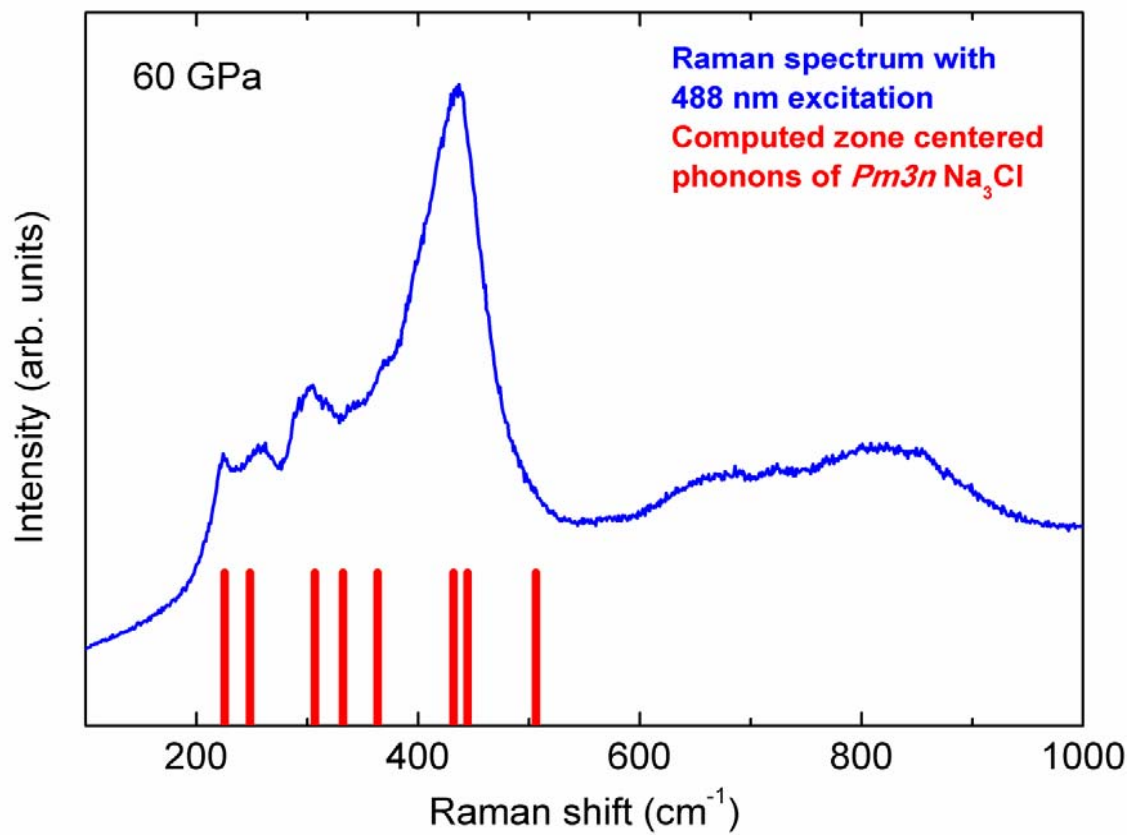

Figure S8. Raman spectrum on sample quenched to 300 K. Vertical bars correspond to the positions of the Brillouin-zone-center optical phonons of the cubic *Pm3n*-NaCl$_3$ computed in this work. Broad Raman peaks at 670 and 820 cm$^{-1}$ can be interpreted as second-order scattering (overtones and combination bands).

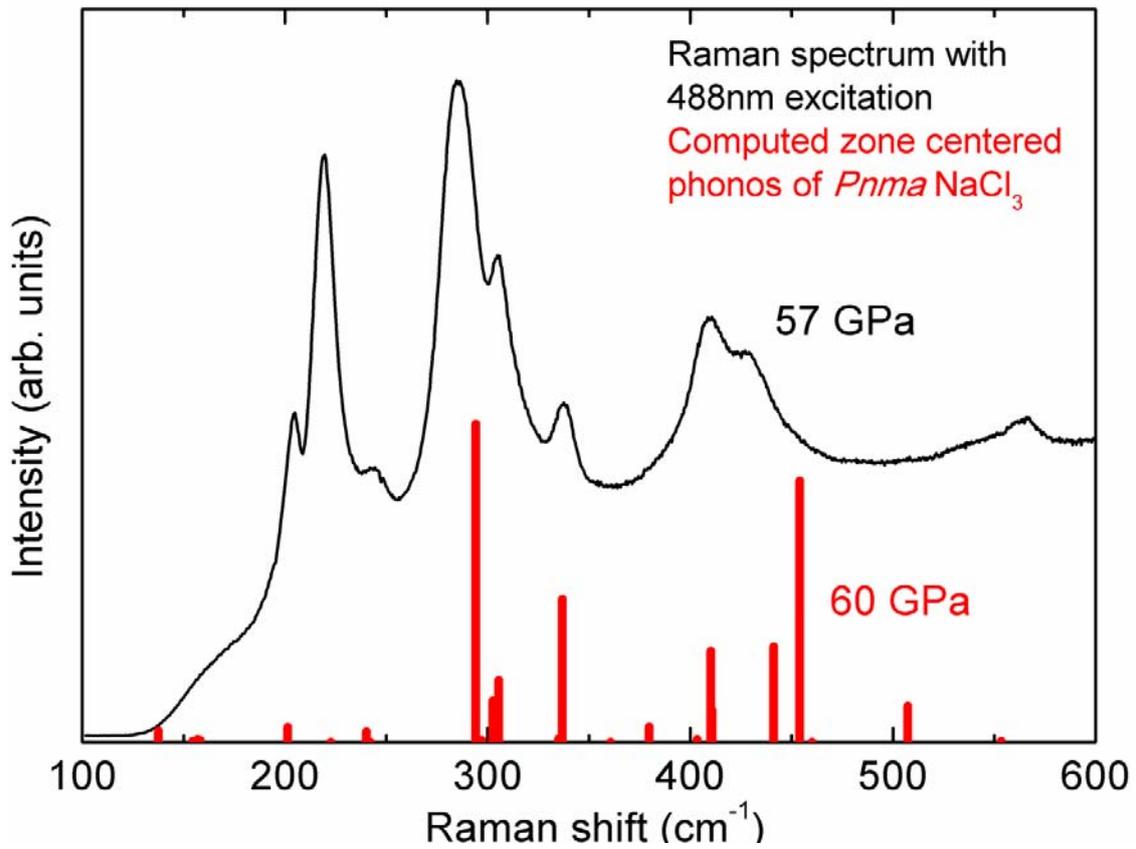

Figure S9. Raman spectrum on sample quenched to 300 K after heating at 60 GPa. Vertical bars correspond to the positions of the Brillouin-zone-center optical phonons of the orthorhombic *Pnma*-NaCl$_3$ computed in this work.

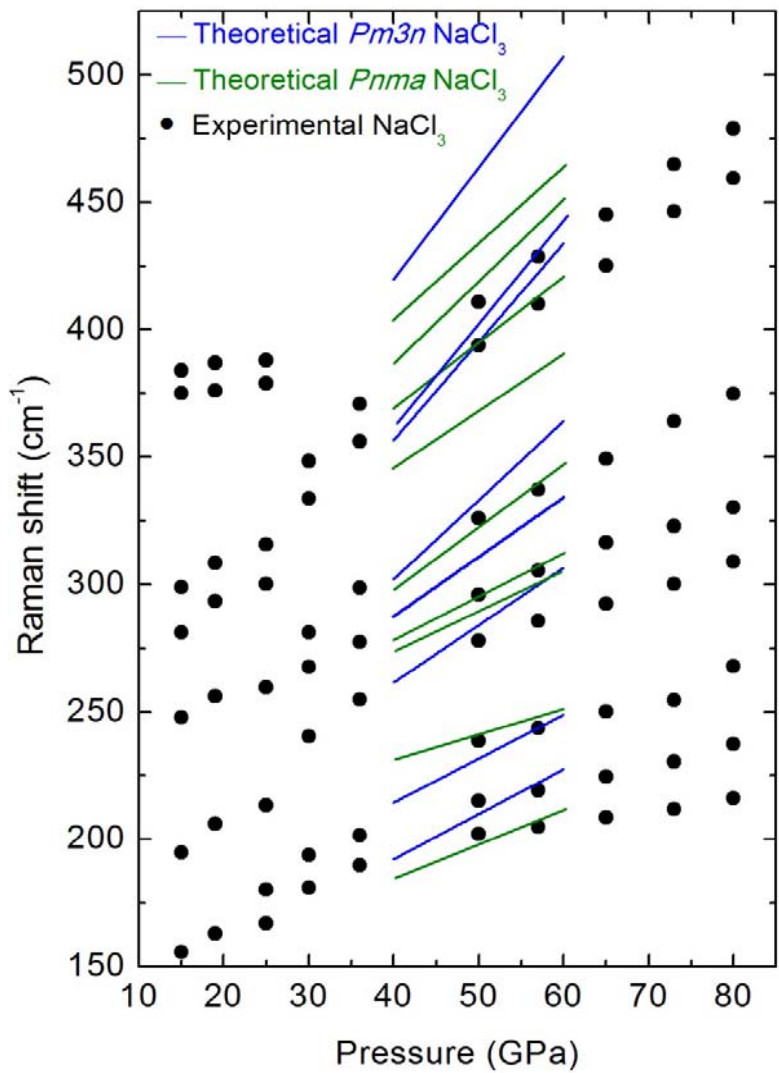

Figure S10. Frequency-pressure plots of the observed Raman modes of the Cl-rich compound (black circles) and calculated zone-center phonons for the cubic (blue lines) and orthorhombic (green lines) phases of $NaCl_3$. The solid lines are the least-squares fits.

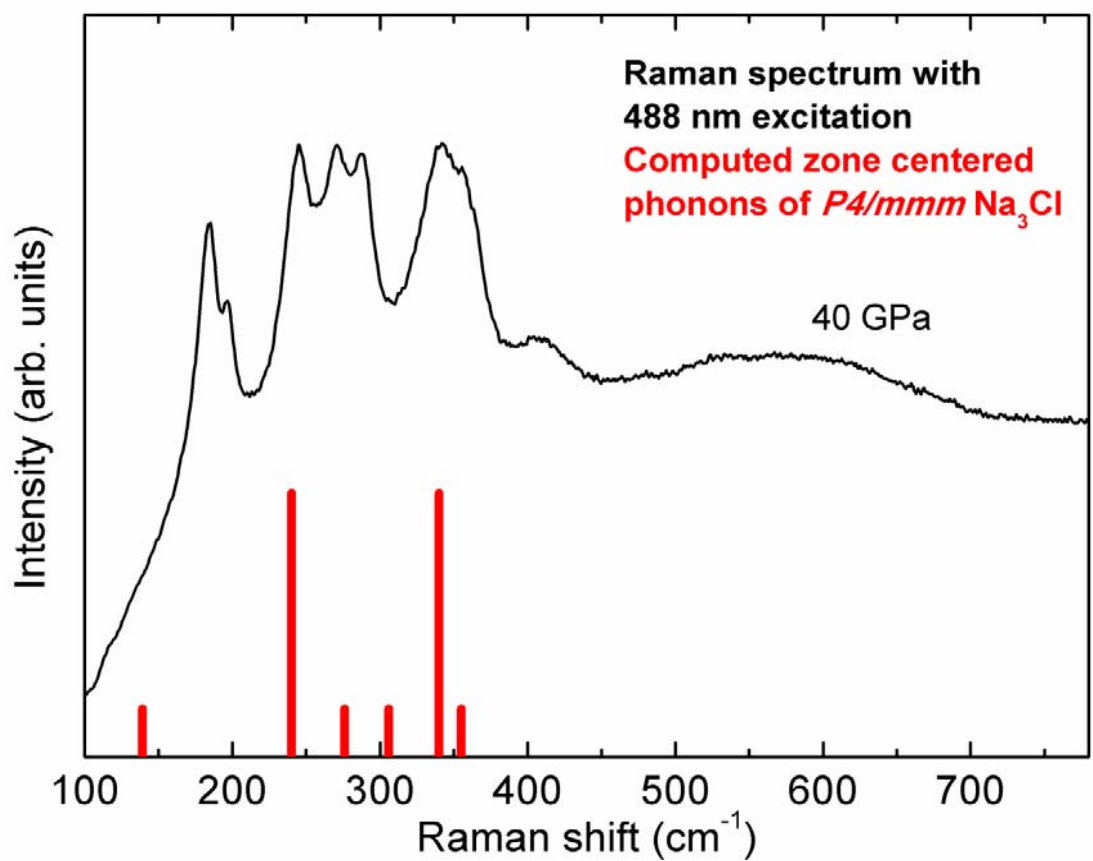

Figure S11. Raman spectrum on the Na-rich material quenched to 300 K and unloaded to 40 GPa. Vertical bars correspond to the positions of the Brillouin-zone-center optical phonons of the tetragonal $P4/mmm$-$Na_3Cl$ computed in this work; two taller bars correspond to Raman active modes.

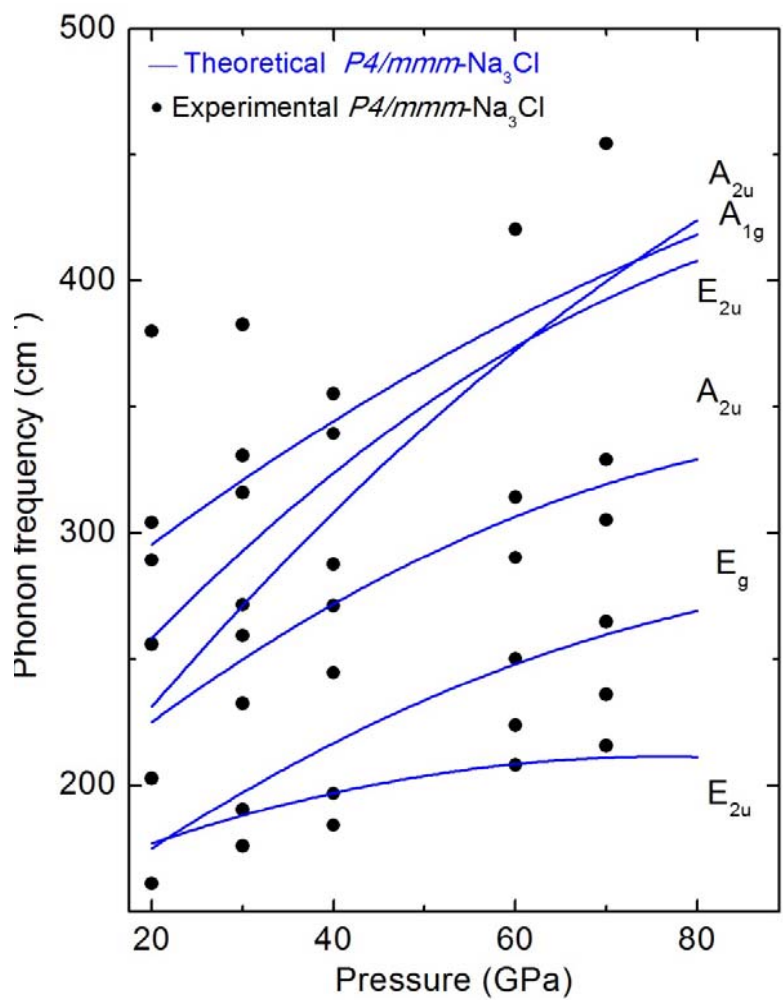

Figure S12. Frequency-pressure plots of observed Raman modes of $Na_3Cl$ (black circles) and calculated zone-center phonons (blue lines).

**Table S1.** Structures of *Pnma*-NaCl$_3$ at 40 GPa, and A15-type (*Pm3n*) NaCl$_3$ and NaCl$_7$ at 200 GPa, and the corresponding atomic Bader charges (Q) and volumes (V).

| | Lattice parameters | Wyckoff position | x | Y | Z | Q, \|e\| | V, Å$^3$ |
|---|---|---|---|---|---|---|---|
| NaCl$_7$ | $a = 4.133$ Å | Na(1a) | 0.000 | 0.000 | 0.000 | +0.827 | 4.30 |
| | | Cl(1b) | 0.500 | 0.500 | 0.500 | +0.067 | 8.63 |
| | | Cl(6g) | 0.248 | 0.500 | 0.000 | -0.149 | 9.61 |
| *Pm3n*-NaCl$_3$ | $a = 4.114$ Å | Na(2a) | 0.000 | 0.000 | 0.000 | +0.823 | 4.16 |
| | | Cl(6d) | 0.250 | 0.500 | 0.000 | -0.275 | 9.90 |
| *Pnma*-NaCl$_3$ | $a = 7.497$ Å $b = 4.539$ Å $c = 6.510$ Å | Na(4c) | 0.342 | 0.250 | 0.480 | +0.833 | 6.77 |
| | | Cl(4c) | 0.141 | 0.250 | 0.180 | -0.524 | 17.67 |
| | | Cl(4c) | 0.383 | 0.250 | 0.966 | -0.035 | 14.67 |
| | | Cl(4c) | 0.095 | 0.250 | 0.741 | -0.274 | 16.27 |

**Typical input file for prediction of stable compounds in the Na-Cl system (USPEX code):**

PARAMETERS EVOLUTIONARY ALGORITHM

*****************************************

*       TYPE OF RUN AND SYSTEM            *

*****************************************

USPEX : calculationMethod (USPEX, VCNEB, META)

301    : calculationType (dimension: 0-3; molecule: 0/1; varcomp: 0/1)

1      : optType (1=enthalpy, 2=volume, 3=hardness, etc.)

% numSpecices

1 0

0 1

% EndNumSpecices

% atomType

Na Cl

% EndAtomType

% valences

1 1

% endValences

*****************************************

*            POPULATION                   *

*****************************************

60     : populationSize (how many individuals per generation)

120    : initialPopSize

50     : numGenerations (how many generations shall be calculated)

20     : stopCrit

*****************************************

%      FIRST VAR. COMP. GENERATION        *

*****************************************

11     : firstGeneMax (how many different compositions for first generation)

8        : minAt (minimum amount of atoms/cell for first generation)

16       : maxAt (maximum amount of atoms/cell for first generation)

*****************************************

*   SURVIVAL OF THE FITTEST AND SELECTION *

*****************************************

0        : reoptOld

0.8      : bestFrac

*****************************************

*           VARIATION OPERATORS           *

*****************************************

0.60   : fracGene (fraction of generation produced by heredity)

0.10   : fracTrans (fraction of the generation produced by permutations)

*****************************************

*              CONSTRAINTS                *

*****************************************

1.6    : minVectorLength ( minimal length of any lattice vector)

% IonDistances

1.0 1.0

0.0 1.0

% EndDistances

*****************************************

*                  CELL                   *

*****************************************

% Latticevalues (this word MUST stay here, type values below)

8.7 8.9

% Endvalues (this word MUST stay here)

*****************************************

*   DETAILS OF AB INITIO CALCULATIONS     *

*****************************************

% supported:   1-vasp, 2-siesta, 3-gulp, 4-LAMMPS, 5-NeuralNetworks

% 6-NeuralNetworks, 7-cp2k, 8-QuantumEspresso, 9-ASE, 10-ATK, 11-CASTEP

abinitioCode (which code from CommandExecutable shall be used for calculation? )

1 1 1 1 1

ENDabinit

%Resolution for KPOINTS - one number per step or just one number in total)

% KresolStart

0.14 0.12 0.10 0.08 0.05

%   Kresolend

% numProcessors (how many processors per calculation)

8 8 8 8 8

% EndProcessors

20     : numParallelCalcs (how many parallel calculations shall be performed)

% commandExecutable

mpirun -np 8 vasp > log

% EndExecutable

*****************************************

*           HARDWARE-RELATED           *

*****************************************

QSH     : whichCluster (default 0, others: CFN, QSH, xservDE)

*****************************************

%           FINGERPRINTS SETTINGS        *

*****************************************

0.03   : sigmaFing

0.08   : deltaFing

10.0   : RmaxFing

0.010 : toleranceFing (tolerance for identical structures)